\documentclass[11pt]{revtex4-2}
\usepackage{graphicx}
 \usepackage{latexsym}
\usepackage{color}
\usepackage{epstopdf}

\usepackage{hyperref}

\newcommand{\be}{\begin{eqnarray}}
\newcommand{\ee}{\end{eqnarray}}

\begin{document}

\title{ Rotating Lee-Wick Black Hole and   Thermodynamics}
\author{Dharm Veer Singh$^a$}
\email{veerdsingh@gmail.com}
\author{Sudhaker Upadhyay$^{b,c,d,e}$}
\email{sudhakerupadhyay@gmail.com; sudhaker@associates.iucaa.in}
\author{Md Sabir Ali$^f$}
\email{sabir@ctp-jamia.res.in}

\affiliation{$^a$Department of Physics, GLA University, Mathura 281406 India}

\affiliation{$^b$Department of Physics, K. L. S. College, Nawada, Bihar 805110, India}
\affiliation{$^c$Department of Physics, Magadh University, Bodh Gaya,
 Bihar  824234, India}
\affiliation{$^d$Inter-University Centre for Astronomy and Astrophysics (IUCAA), Pune, Maharashtra 411007, India}
\affiliation{$^e$School of Physics, Damghan University, P.O. Box 3671641167, 
Damghan, Iran}
\affiliation{$^f$Center for Theoretical Physics, Jamia Millia Islamia, New Delhi, India}

\begin{abstract}
We derive a singular solution for the rotating counterpart of Lee-Wick gravity having a point  source in a higher-derivative theory. We 
critically analyze the thermodynamics of such a thermal system by  evaluating mass parameters, angular velocity, and Hawking temperature. The system follows the first law of thermodynamics and leads to the expression of entropy.
We further discuss the stability and phase transition of the theory by evaluating heat capacity and free energy. The phase transition occurs at the point of divergence and the temperature is maximum.
Remarkably, the black hole is unstable for a small horizon radius and stable for a large horizon radius.
\end{abstract}

\maketitle


\section{Introduction \label{s-1}}

Quantizing the theory of gravity is one of the most challenging tasks of  theoretical physics. It is known that the  gravity theory with forth derivative terms is renormalizable   which in fact violate the unitarity of the theory \cite{1}.  However,  the  fourth derivative terms lead to the negative-norm states  in the physical spectrum of the theory. This justifies instabilities in the classical solutions.  A consistent theory of quantum gravity theory which  agrees with both unitarity and renormalizability requires a modification in 
Einstein theory of gravity. One of the  possible solutions to this problem    is to consider a local gravitational theory with higher 
than fourth derivative terms. Such theories with higher  derivatives terms have very interesting  properties as they are superrenormalizable \cite{2}.  Also, these theories are unitary  in the Lee-Wick formalism for the   massive complex poles  \cite{3,4}. Hence, these theories remove  the conflict between unitarity and renormalizability  in quantum gravity  \cite{w1,w2,w3,l1,l2,l3,l4}.   Also, the proof of unitarity of these models do not require any  fine-tuning   and therefore prove themselves as a better alternative  to the non-local   theories \cite{5,6,7,8,9}.

The singular solutions (black holes) are the most fundamental objects in 
gravity theory as they provide powerful probes to the study of various   aspects of the theory.  Therefore, the search of black-hole 
solutions in the    higher-derivative gravity theories are of considerable  interests. A static vacuum solution of an approximate equation of motion   of the Lee-Wick gravity is studied recently  where the solution is  generated by  a point-like massive source \cite{Bambi:2016wmo}. The authors of Ref.  \cite{Bambi:2016wmo} discussed  different scenarios by comparing the mass of  the source with a critical value. They also shed light on the thermodynamics   of these black holes and their evaporation process.  The thermodynamics  of the black hole plays a vital role in   understanding the quantum theory of gravity \cite{10,11,12,13,14}.   Hawking and Page  had found that the black hole solutions in asymptotically $AdS$ space have thermodynamic properties  \cite{haw}. The rotational counterpart of these gravity theory remains unexplored. This provides us an opportunity to generalize the work and  this is the motivation of present study. 
 
 In this paper, we obtain a rotating  Lee-Wick black hole solution by using a modified Newman-Janis algorithm \cite{Newman:1965tw,Hansen:2013owa,az,ze}. To do so, we first convert static spherically symmetric Lee-Wick black hole solution into its rotational counterpart, we transform the coordinates of the Lee-Wick metric  from the Boyer-Lindquist  to the Eddington-Finkelstein.  The numerical analysis  leads to the possibilities to find non-vanishing values of rotation parameter which corresponds  minimum   metric function  \cite{Ahmed:2020dzj,Ahmed:2020ifa}.  
We also investigate the basic thermodynamics properties of rotating Lee-Wick  black hole. We first derive mass parameter and angular velocity. Moreover, we  compute the Hawking temperature of rotating lee-Wick black hole  in terms of  temperature of Kerr black hole.  A comparative analysis  suggests that rotating Lee-Wick black holes with small rotation parameter are  comparatively   hotter. Also, the rotation parameter plays more significant role for small black holes. In order to study the stability of such black holes, we need to calculate heat capacity of the black hole and check the sign of heat capacity and free energy. Here, we find three phase transitions. For the stationary case when rotation parameter is switched-off,  the phase transition occurs only once. In contrast, when rotation parameter acquires non-zero values, the phase transition occurs. Also,  in stationary case, the larger black holes are stable.

The paper is organized as follows. In section \ref{sec2}, we obtain a  rotational counterpart of Lee-Wick black hole solution. In section 
\ref{sec3}, we discussed the thermodynamics of rotational Lee-Wick black hole and studies their local and global stability. Finally, we conclude the work in the section \ref{sec4}.

\section{Generating rotating Lee-Wick black hole solutions}\label{sec2}.
  We start with the action of the minimal theory having  unitarity   with the Lee-Wick prescription
 \cite{Bambi:2016wmo}
\begin{equation}
S=\frac{1}{8\pi G_N}\int d^4 x\sqrt{-g} \left[R+ {\Lambda^{-4}} G_{\mu\nu}\Box R^{\mu\nu}\right],\label{1}
\label{action}
\end{equation}
where   $R$ is the Ricci scalar, $\Lambda$  represents the UV scale of the theory and $G_{\mu\nu}$ is the Einstein tensor.  Varying the action (\ref{action}), we obtain the equations of motion (EoM)
\begin{equation}
\left(1+\frac{\Box^2}{\Lambda^4}\right)G_{\mu\nu}+O(R^2_{\mu\nu})=8\,\pi \,G_N T_{\mu\nu}
\end{equation}
where energy momentum tensor (EMT)  $T_{\mu\nu}= \left(1+{\Box^2}/{\Lambda^4}\right){\tilde T}_{\mu\nu}$. Ignoring higher-order ($O(R^2_{\mu\nu})$) terms, this EoM takes following form \cite{Bambi:2016wmo}:
\begin{equation}
G_{\mu\nu}=8\,\pi \, G\,\left(1+\frac{\Box^2}{\Lambda^4}\right)^{-1}{ T}_{\mu\nu}.\label{3}
\end{equation}
The only non-vanishing component of  EMT  is $T_t^t=-M\delta(x)$ and, for this , the effective EMT is given by
\begin{equation}
{ \tilde T}^{\mu}_{\nu}=\text{diag}\left(-{\tilde \rho},{\tilde P_r},{\tilde P_{\theta}},{\tilde P_{\theta}}\right),
\label{emt}
\end{equation}
where ${\tilde \rho}$, ${\tilde P_r}$ and ${\tilde P_{\theta}}$ are the effective energy density, effective radial pressure and effective tengential pressure, repectively. The effective energy density is calculated as \cite{Bambi:2016wmo}
\begin{eqnarray}
\tilde \rho=\left(1+\frac{\Box^2}{\Lambda^4}\right)^{-1}&=&-M\delta(x)=M\int \frac{dk^3}{(2\pi)^3} \frac{e^{ikx}}{1+k^4/\Lambda^4}\nonumber\\
&=&\frac{M\Lambda^2}{4\pi r}\sin \frac{r\Lambda}{\sqrt{2}}e^{-\frac{r\Lambda}{\sqrt{2}}}.
\end{eqnarray}

Now, we write  the static and spherically symmetric Lee-Wick solution 
which  matches with the form of Schwarzschild metric as follows 
\cite{Bambi:2016wmo},
\be\label{metric}
ds^2 = -f(r)dt^2 + \frac{dr^2}{f(r)} + r^2 d\Omega_2^2 \quad \text{with}\quad  
f(r) = 1- \frac{2 m(r)}{r}.
\ee
Here $m(r)$ refers to effective mass and depends on the radial coordinate ($r$) 
only due to the spherical symmetry. The expression for mass function $m(r)$ 
is given by \cite{Bambi:2016wmo}
\begin{eqnarray}
m(r)=M\left[1-e^{-\frac{\Lambda r}{\sqrt{2}}}\left(1+\frac{\Lambda r}
{\sqrt{2}}\right)\cos\left(\frac{\Lambda r}{\sqrt{2}}\right)+\frac{\Lambda r}
{\sqrt{2}}\sin\left(\frac{\Lambda r}{\sqrt{2}}\right)\right],
\end{eqnarray}
where $\Lambda$  is the UV scale  and $M$ corresponds to the 
mass of a static point-like source. It should be noted  that   this is not an exact solution
of the equations of motion of the model (\ref{1}) but rather a solution to the `approximate' equations of motion (2)  in \cite{Bambi:2016wmo} when the quadratic and higher-order terms in the Ricci
tensor are neglected.

{ Upon   effective mass expansion around the centre ($r=0$), one finds de Sitter 
core  \cite{Bambi:2016wmo} 
\begin{eqnarray}\label{metric-ds1}
ds^2\approx-\left(1-\frac{\Lambda_{eff}r^2}{3}\right)dt^2+\frac{1}{\left(1-
\frac{\Lambda_{eff}r^2}{3}\right)}dr^2+r^2d\Omega^2_{2}
\end{eqnarray} }
where $\Lambda_{eff}=\sqrt{2}\Lambda^3 M$ is the effective cosmological 
constant. The metric (\ref{metric-ds1}) describes a regular spacetime and 
thus curvature is singularity free. 

\begin{figure*}[ht]
\begin{tabular}{c c c c}
\includegraphics[width=.75\linewidth]{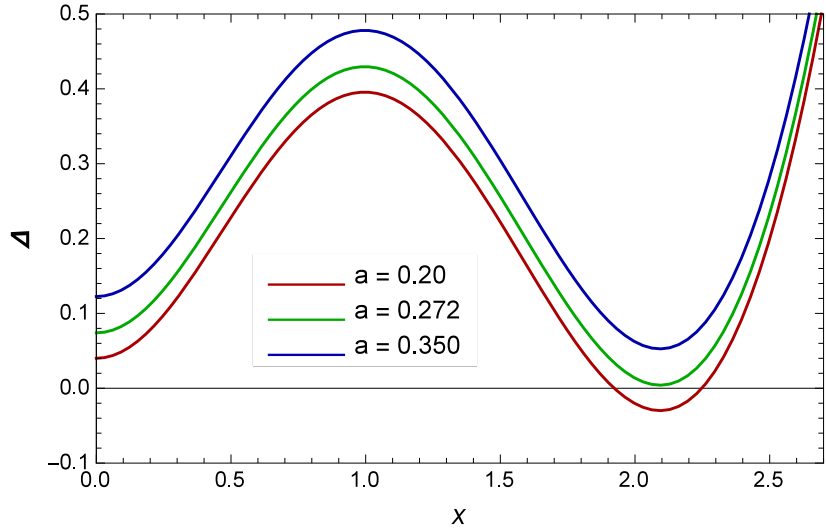}
\end{tabular}
\caption{The plot of metric function $\Delta(r)$ as the function of dimensionless parameter $x=\Lambda r/\sqrt{2}$ for different values of angular momentum   $a$  with fixed value of  $\Lambda=4\sqrt{2}\pi$. }
\label{fr1}
\end{figure*}

 Now, we proceed to obtain the rotating counterpart of the metric 
(\ref{metric}). In order to do so, we follow the method 
 proposed originally by Newman and Janis \cite{Newman:1965tw} and further modified by  
Azreg-Ainou~ \cite{az}. In order to convert static spherically symmetric Lee-Wick black hole solution into 
its rotational counterpart, {  we first transform the coordinates of the spacetime metric~(\ref{metric}) 
from the Boyer-Lindquist    ($t,r,\theta,\phi$) to the Eddington-Finkelstein    ($u,r,\theta,\phi$)  as following:
\begin{eqnarray}\label{2.1}
du=dt-\frac{dr}{f(r)}.
\end{eqnarray}
This leads to  the  final spacetime metric in the following form:
\begin{eqnarray}\label{2.2}
ds^2=-fdu^2-2dudr+r^2 d\theta^2+r^2\sin^2\theta d\phi^2.
\end{eqnarray}
It is known that contravariant components of the metric tensor in the advanced null Eddington-Finkelstein coordinates can be expressed by the null tetrad of the form
\begin{eqnarray}\label{2.3}
g^{\mu\nu}=-l^\mu n^\nu-l^\nu n^\mu+m^\mu\bar{m}^\nu+m^\nu\bar{m}^\mu,
\end{eqnarray}
Following the method as discussed in Ref. \cite{ze},  we first write the 
metric coordinates  in null tetrad and do  complex coordinate transformations in the $u-r$ plane as
\begin{eqnarray}\label{2.5}
u\rightarrow u-ia\cos\theta, \quad r\rightarrow r-ia\cos\theta.
\end{eqnarray}
This converts the metric functions  into a new form: $f(r)\rightarrow F(r,a,
\theta)$, $r^2\rightarrow \Sigma(r,a,\theta)$.  Furthermore, null tetrads also take the following new form:
\begin{eqnarray}\label{2.6}
&&l^\mu=\delta_r^\mu\ , \quad m^\mu=\frac{1}{\sqrt{2\Sigma}}\left[\delta_\theta^\mu+ia\sin\theta(\delta_u^\mu-\delta_r^\mu)+\frac{i}{\sin\theta}\delta_\phi^\mu\right]\ ,\nonumber\\
&&n^\mu=\sqrt{\frac{G}{F}}\delta_u^\mu-\frac{1}{2}F\delta_r^\mu\ , \quad \bar{m}^\mu=\frac{1}{\sqrt{2\Sigma}}\left[\delta_\theta^\mu-ia\sin\theta(\delta_u^\mu-\delta_r^\mu)-\frac{i}{\sin\theta}\delta_\phi^\mu\right].
\end{eqnarray}
Then we can rewrite the contravariant components of the metric tensor $g^{\mu\nu}$ by using~(\ref{2.3}) as
\begin{eqnarray} \label{2.7}
g^{\mu\nu} =      \left(
\begin{array}{c c c c}
a^2\sin^2\theta/\Sigma & -\sqrt{G/F}-a^2\sin^2\theta/\Sigma & 0 &a/\Sigma \\
-\sqrt{G/F}-a^2\sin^2\theta/\Sigma & G+a^2\sin^2\theta/\Sigma & 0 & -a/\Sigma \\
0 & 0 & 1/\Sigma & 0 \\
a/\Sigma & -a/\Sigma & 0 & 1/(\Sigma\sin^2\theta) \\
\end{array} \right).
\end{eqnarray}
The covariant components read
\begin{eqnarray} \label{2.8}
g_{\mu\nu} =      \left(
\begin{array}{c c c c}
-F & -\sqrt{\frac{F}{G}} & 0 &a\left(F-\sqrt{\frac{F}{G}}\right)\sin^2\theta \\
-\sqrt{\frac{F}{G}} & 0 & 0 & a\sqrt{\frac{F}{G}}\sin^2\theta \\
0 & 0 & \Sigma & 0 \\
a\left(F-\sqrt{\frac{F}{G}}\right)\sin^2\theta & a\sqrt{\frac{F}{G}}\sin^2\theta & 0 & \sin^2\theta\left[\Sigma+a^2\left(2\sqrt{\frac{F}{G}}-F\right)\sin^2\theta\right] \\
\end{array} \right).
\end{eqnarray}
The last step of the algorithm is to turn back from the Eddington-Finkelstein  coordinates to the Boyer-Lindquist  coordinates by using the following coordinate transformations:
\begin{eqnarray}\label{2.9}
du=dt+\lambda(r)dr\  \qquad\text{and}\qquad \quad d\phi=d\phi+\chi(r)dr\ .
\end{eqnarray}
The transformation functions $\lambda(r)$ and $\chi(r)$ are found due to the requirement that, except the coefficient $g_{t\phi}$ ($g_{\phi t}$), all the non-diagonal components of the metric tensor  are equal to zero. Thus
\begin{eqnarray}\label{2.10}
&&\lambda(r)=-\frac{r^2+a^2}{f(r)^2+a^2}\ ,\qquad\qquad \qquad\qquad\chi(r)=-\frac{a}{f(r)r^2+a^2},\nonumber\\
&&F(r,\theta)=\frac{(f^2+a^2\cos^2\theta)\Sigma}{(r^2+a^2\cos^2\theta)^2},\qquad\text{and}\qquad G(r,\theta)=\frac{f^2+a^2\cos^2\theta} {\Sigma}.
\end{eqnarray}
 }
Consequently,  we obtain the following line-element for the  rotational  counterpart of the Lee-Wick black hole solution
\begin{eqnarray}\label{metric-ds2}
ds^2&=&-\left(1-\frac{2Mr-2Me^{-\frac{\Lambda r}{\sqrt{2}}}\left[\left(1+
\frac{\Lambda r}{\sqrt{2}}\right)\cos\frac{\Lambda r}{\sqrt{2}}+\sin
\frac{\Lambda r}{\sqrt{2}}\right]}{r^2+a^2 \cos^2\theta}\right)dt^2+\frac{r^2+a^2 \cos^2\theta}{\Delta}
dr^2\nonumber\\
&-&2a\sin^2\theta\left(\frac{2Mr-2Me^{-\frac{\Lambda r}{\sqrt{2}}}\left[\left(1+\frac{\Lambda r}{\sqrt{2}}\right)\cos\frac{\Lambda r}{\sqrt{2}}+\sin\frac{\Lambda r}{\sqrt{2}}\right]}{r^2+a^2 \cos^2\theta}\right)d\phi dt+(r^2+a^2 \cos^2\theta) d\theta^2 \nonumber\\
& +&\sin^2\theta\left[r^2+a^2+a^2\sin^2\theta\left(\frac{2Mr-2Me^{-\frac{\Lambda r}{\sqrt{2}}}\left[\left(1+\frac{\Lambda r}{\sqrt{2}}\right)\cos\frac{\Lambda r}{\sqrt{2}}+\sin\frac{\Lambda r}{\sqrt{2}}\right]}{r^2+a^2 \cos^2\theta}\right)\right]d\phi^2, 
\label{lw}
\end{eqnarray}
with
\begin{eqnarray}\label{2.14}
\Delta=r^2-2Mr+a^2+2Me^{-\frac{\Lambda r}{\sqrt{2}}}\left[\left(1+\frac{\Lambda r}{\sqrt{2}}\right)\cos\frac{\Lambda r}{\sqrt{2}}+\sin\frac{\Lambda r}{\sqrt{2}}\right],
\end{eqnarray}
where $a$ is the rotation parameter of black hole. Arising from the  UV scale theory, $\Lambda$  measures the potential deviation from the Kerr metric. The obtained  black hole solution (\ref{lw}) is independent of $t, \phi$, which implies that it admits two Killing vectors given by $\eta^{\mu}=\delta^{\mu}_t$ and $\xi^{\mu}=\delta^{\mu}_{\phi}$. The horizons of rotating Lee-Wick black holes are solution of equation  $(\eta^{\mu}\xi_{\mu})^2-(\eta_{\mu}\eta^{\mu})(\xi_{\nu}\xi^{\nu})=0$, which gives
\be
g_{t\phi}^2-g_{tt}g_{\phi\phi}=0.
\ee
Here, this yields
\be
r^2-2Mr+a^2+2Me^{-\frac{\Lambda r}{\sqrt{2}}}\left[\left(1+\frac{\Lambda r}{\sqrt{2}}\right)\cos\frac{\Lambda r}{\sqrt{2}}+\sin\frac{\Lambda r}{\sqrt{2}}\right]=0.
\ee
This is a transcendental  equation which cannot be solved analytically.  The numerical analysis of the $\Delta=0$ on  varying the angular momentum  $a$ with fixed mass $M=1$ is depicted in the Fig. \ref{fr1}. The numerical analysis of $\Delta = 0$  reveals that it is possible to find non-vanishing value of angular momentum ($a$) and  UV scale parameter $x=\Lambda r/\sqrt{2}$  for which metric function $\Delta$ is minimum, i.e, $\Delta(x_+ ) = 0$ and this will give two real roots $x_+$ and $x_-$which  correspond to the Cauchy horizon and event horizon.

\begin{table}[ht]
 \begin{center}
 \begin{tabular}{ l | l   | l   | l      }
\hline
            \hline
  \multicolumn{1}{c|}{ $a$} &\multicolumn{1}{c}{$x_+$}  &\multicolumn{1}{|c|}{$x_-$}  &\multicolumn{1}{c}{$\delta=x_+-x_-$} \\
            \hline
            \,\,\,\,\,0.10 ~~  &~~1.85~~  & ~~2.31.~~ & ~~0.46~  \\    
            \,\,\,\,\,0.20~~ &~~1.92~~ & ~~2.24~~ & ~~0.32~~    \\
 \,\,\,\,\,0.272~~ &~~2.096~~ & ~~2.096~~ & ~~0~~    \\
            \hline 
\hline
        \end{tabular}
        \caption{The table for horizon radius for different value of rotation parameter $a$   with fixed value of  $\Lambda=4\sqrt{2}\pi$.}
\label{tr10}
    \end{center}
\end{table}
{ From the Fig. (\ref{fr1}) and Table \ref{tr10}, we can see that the horizon of the black hole decreases with increasing   rotation parameter and at $a=0.268$  two horizons reduces to one, i.e., $r_{\pm}= r_E$ such that $\Delta(r_E)=\Delta'{r_E}=0$.} 

\section{Thermodynamics}\label{sec3}
In this section, we analyse the thermodynamics of the rotating Lee-Wick black hole described by metric ~(\ref{metric-ds2}). Let us begin by deriving the mass by $g^{rr}=0$  as following:
\begin{eqnarray}
M_+=\frac{\left(a^2+r_+^2\right)}{2r_+ \left(1-e^{-\frac{\Lambda r_+}{\sqrt{2}}}\left(\left(1+\frac{\Lambda r_+}{\sqrt{2}}\right)\cos\left[\frac{\Lambda r_+}{\sqrt{2}}\right]+\frac{\Lambda r_+}{\sqrt{2}}  \sin\left[\frac{\Lambda r_+}{\sqrt{2}}\right]\right)\right)}.
\label{mass}
\end{eqnarray}
In the absence of rotation parameter ($a$), the mass (\ref{mass}) reduces to the mass of Lee-Wick black hole \cite{Bambi:2016wmo}; however this reduces to mass of  Kerr black hole in the limit of $\Lambda\to 0$.

For the stationary axially symmetric metric like (\ref{metric-ds2}), we have two associated Killing vectors   $\eta^{\mu}=\partial_t$ corresponding to the time translational symmetry along the $t$-axis and $\xi^{\mu}=\partial_\phi$ corresponding to the rotational symmetry about $\phi$-axis. In therms of these killing vectors a Killing field $\chi{^\mu}$ can be expressed as $\chi{^\mu}=\eta^{\mu}+\Omega \xi^{\mu}$, where $\Omega$ refers to the angular velocity of the metric. Being  a null vector at the event horizon (i.e. $\chi{^\mu}\chi_{\mu}=0$),  the Killing field decides the angular velocity   and leads to \cite{Ali:2019rjn,Ali:2019myr}
\begin{eqnarray}
g_{tt}+2\Omega g_{t\phi}+\Omega^2 g_{\phi\phi}=0,
\end{eqnarray}
where angular velocity is given by
\begin{eqnarray}
\Omega &=&-\frac{g_{t\phi}}{g_{\phi\phi}}\pm\sqrt{\left(\frac{g_{t\phi}}{g_{\phi\phi}}\right)^2-\frac{g_{tt}}{g_{\phi\phi}}},\nonumber\\
&=&\frac{-a\left[\Sigma-(\Delta-a^2\sin^2\theta)\right]}{\Sigma^2+a^2\sin^2\theta\left[2\Sigma-(\Delta-a^2\sin^2\theta)^2\right]}
\nonumber\\
&\pm &\frac{\Sigma\Delta^{\frac{1}{2}}}{\sin\theta\left[(a^2\sin^2\theta+\Sigma)^2-\Delta a^2\sin^2\theta\right]}.
\end{eqnarray} 
In terms of horizon radius, the angular velocity reads
\begin{eqnarray}\label{omega}
\Omega_+=-\left.\frac{g_{t\phi}}{g_{\phi\phi}}\right|_{r=r_+}=\frac{a}{r_+^2+a^2}.
\end{eqnarray}
Before computing  temperature of the black holes, it is worth to first work out the surface gravity at the event horizon
 for the obtained solution (\ref{metric-ds2}). This gives  
\begin{eqnarray}\label{kappa}
\kappa=\sqrt{-\frac{\nabla_{\mu}\chi{_\nu}\nabla{^\mu}\chi{^\nu}}{2}}=\frac{\Delta^\prime(r_+)}{2(r_+^2+a^2)}.
\end{eqnarray}
Here ${}^\prime$ denotes differentiation with respect to horizon radius.
Utilizing the standard relation of Surface gravity  and Hawking temperature $T_+=\kappa/2\pi$,  we obtain the Hawking temperature of the rotating Lee-Wick black hole metric (\ref{metric-ds2}) as
\begin{eqnarray}\label{temp1}
T_+=T_+^{Kerr}\left[\frac{e^{-\frac{r_+ \Lambda }{\sqrt{2}}}\left(\left(1+ \frac{\Lambda r_+ }{\sqrt{2}}\right) \cos\left[\frac{ \Lambda r_+}{\sqrt{2}}\right]+r_+ \left( \frac{\Lambda r_+ }{\sqrt{2}}+\frac{\Lambda }{2}\left(\frac{a^2+r_+^2}{a^2-r_+^2} \right)\right) \sin\left[\frac{ \Lambda r_+}{\sqrt{2}}\right]\right)-1}{e^{-\frac{r_+ \Lambda }{\sqrt{2}}}\left(\left(1+ \frac{\Lambda r_+}{\sqrt{2}}\right) \cos\left[\frac{\Lambda r_+}{\sqrt{2}}\right]+\frac{\Lambda r_+}{\sqrt{2}}\sin\left[\frac{\Lambda r_+}{\sqrt{2}}\right]-r_+\right)}\right],
\label{tlw}
\end{eqnarray}
where $ T_+^{Kerr}={\left(r_+^2-a^2\right)}/{\left(4\pi r_+(r_+^2+a^2)\right)}$ is the temperature of the Kerr black hole. The temperature (\ref{tlw}) reduces to the temperature of the Kerr black hole in the limit of $\Lambda \to 0$.

\begin{figure*}[ht]
\begin{tabular}{c c c c}
\includegraphics[width=9cm, height=8.5cm]{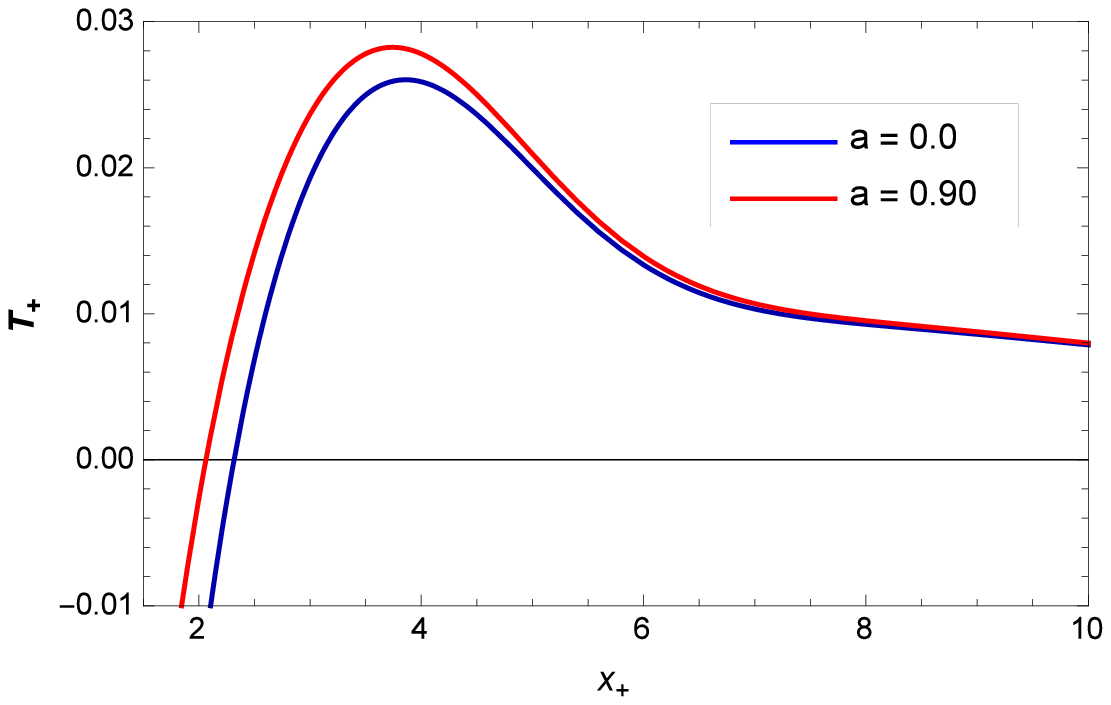}
\includegraphics[width=9cm, height=8.5cm]{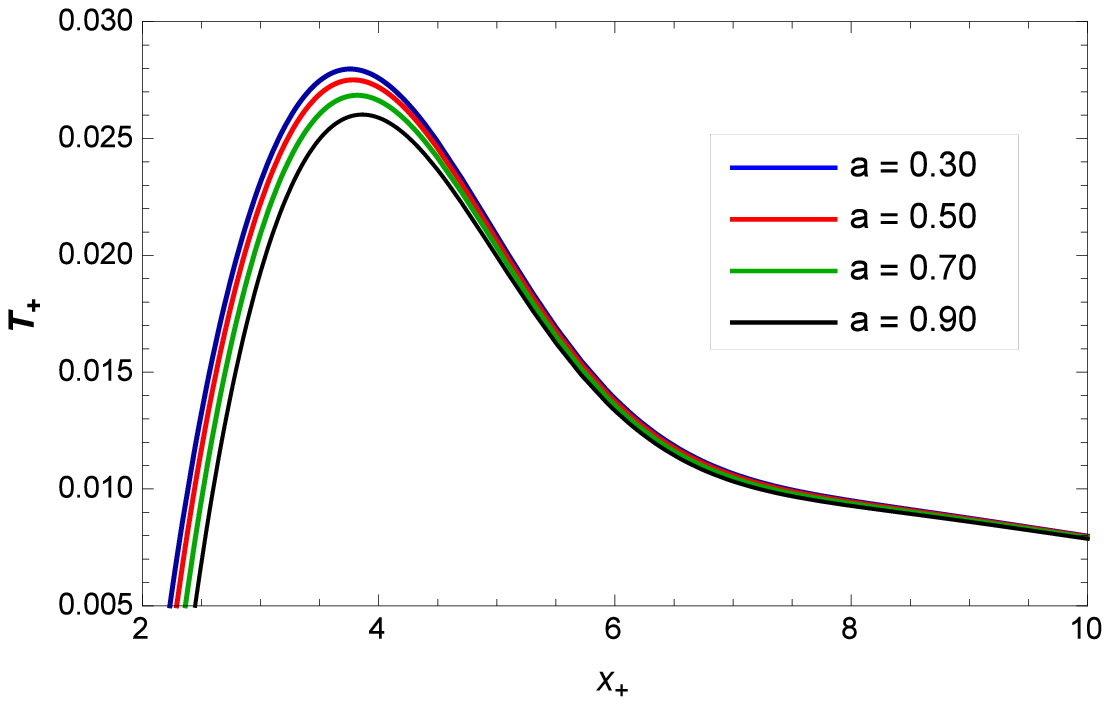}
\end{tabular}
\caption{The plot of Hawking temperature $T_+$ versus horizon radius $x_+$ for different values of rotation parameter  $a$  with  fixed Value of $\Lambda=4\sqrt{2}\pi$. }
\label{fig:6}
\end{figure*}

\begin{table}[ht]
 \begin{center}
 \begin{tabular}{ l | l   | l   | l  | l      }
\hline
            \hline
  \multicolumn{1}{c|}{ $a$} &\multicolumn{1}{c}{$0.30$}  &\multicolumn{1}{|c|}{$0.50$}  &\multicolumn{1}{c|}{$0.70$}&\multicolumn{1}{c}{$0.90$} \\
            \hline
            \,\,\,\,\,$T_{max}$ ~~  &~~0.0279~~  & ~~0.0275.~~ & ~~0.0268~ ~&~~0.0260~ ~ \\    
            \,\,\,\,\,$x_c$~~ &~~3.750~~ & ~~3.797~~ & ~~3.816~~  &~~3.872~ ~  \\
            
            \hline 
\hline
        \end{tabular}
        \caption{The table for temperature for different value of rotation paramete $a$ with fixed value of  $\Lambda=4\sqrt{2}\pi$.}
\label{tr11}
    \end{center}
\end{table}

{ From the Fig. \ref{fig:6} and table (\ref{tr11}), it is clear that the black hole becomes hotter  for small rotation parameter and the critical radius increases   with the rotation parameter $(a)$. The maximum Hawking temperature decreases for higher values of the critical radius $(r_c)$. Thus, This  signifies that  the effect of rotation parameter is  more significant for small black holes. }

\begin{figure*}[ht]
\begin{center}$
\begin{array}{c c c c}
\includegraphics[width=.75\linewidth]{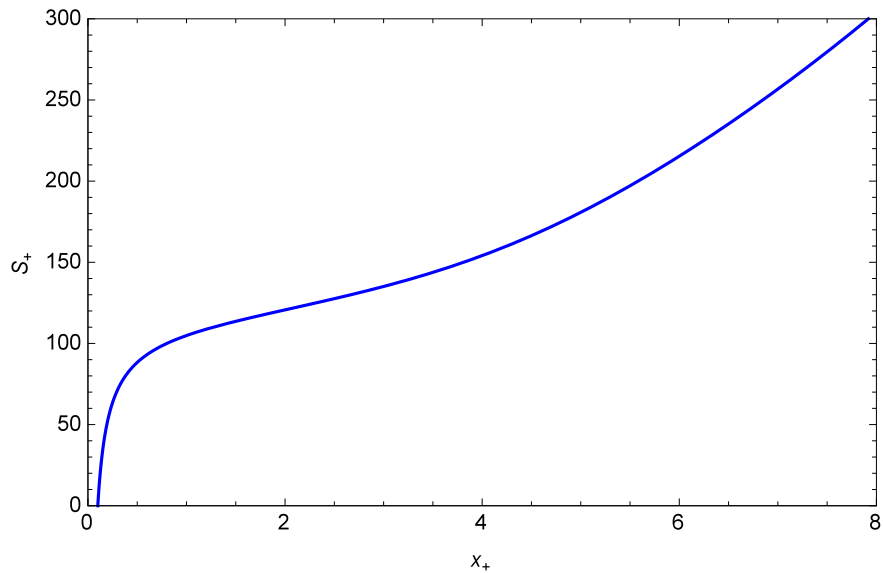}
\end{array}$
 \end{center}
\caption{The plot of entropy $S_+$ vs $x_+$ with fixed value of  $\Lambda=4\sqrt{2}\pi$. }
\label{7}
\end{figure*}

This black hole can be considered as a thermodynamic system only if the   quantities associated with it must satisfies the first-law of 
  thermodynamics 
\begin{eqnarray}\label{1law}
dM_+ = T_+dS_+ + \Omega_+ da.
\end{eqnarray}
The validity of above relation leads  to the expression for entropy 
\begin{equation}
S=-4\pi\int dr_+\frac{(a^2+r_+^2) e^{-r_+\Lambda/\sqrt{2}}}{e^{-\frac{\Lambda r_+}
{\sqrt{2}}}\left(\left(2+{\sqrt{2}}\,r_+ \Lambda\right)\cos\left[\frac{\Lambda 
r_+}{\sqrt{2}}\right]+{\sqrt{2}}  \sin\left[\frac{\Lambda r_+}{\sqrt{2}}
\right]\right)}.
\label{entropy}
\end{equation}
From this this expression it is clear that the entropy does not depends upon roation parameter ($a$). The entropy is the function of horizon radius which is depicted in Fig. (\ref{7})

 The definition of first-law of thermodynamics is worth mentioning to check the validity of these thermodynamic quantities. It is easy to verify the complete form of the first-law of thermodynamics as
\begin{equation}
dM_+=T_+dS_++\Omega_+ dJ
\label{firstlaw}
\end{equation}
Substituting the value of mass, temperature and entropy from the Eqs. (\ref{mass}), (\ref{tlw}) and (\ref{entropy}), respectively, given in Eq. (\ref{firstlaw}),  we can see that the rotating Lee-Wick black hole follows the first-law of thermodynamics. 
\section{Local and Global Stability}
Finally, we  study  the thermodynamic stability of rotating Lee-Wick black hole  by estimating the  heat capacity. The stability of the black hole can be estimated from sign of the heat capacity ($C_+$) as positive heat capacity reflects the stable black hole and negative reflects the unstable one.  The heat capacity of the black hole can be calculated by \cite{Ali:2019rjn,Ali:2019myr,
Singh:2017bwj,Chaturvedi:2016fea}
\begin{eqnarray}
C_+=\frac{dM_+}{dT_+}=\frac{dM_+}{dr_+}\frac{dr_+}{dT_+}
\label{sh1}
\end{eqnarray}
substituting the mass and temperature from Eq. (\ref{mass}) and Eq. (\ref{tlw}) into Eq. (\ref{sh1}) we get
\begin{eqnarray}
C_+=2\pi\frac{ \left(r^2-a^2\right)\left(r^2+a^2\right)^2 Y}{X},
\label{f7}
\end{eqnarray}
with
\begin{eqnarray}
Y&=& e^{-\frac{\Lambda r_+}{\sqrt{2}}}\left(\left(1+ \frac{\Lambda r_+ }{\sqrt{2}}\right) \cos\left[\frac{ \Lambda r_+}{\sqrt{2}}\right]+r_+ \left( \frac{\Lambda r_+ }{\sqrt{2}}+\frac{\Lambda }{2}\left(\frac{a^2+r_+^2}{a^2-r_+^2} \right)\right) \sin\left[\frac{ \Lambda r_+}{\sqrt{2}}\right]\right)-1,\nonumber\\
X&=&2\left(a^4+4a^2r_+^2-r_+^4\right)\left[\left(e^\frac{{\Lambda r_+}}{\sqrt{2}}-\cos\left[\frac{\Lambda r_+}{\sqrt{2}}\right]\right)^2+\frac{{2\Lambda r_+}}{\sqrt{2}}\left(e^\frac{{\Lambda r_+}}{\sqrt{2}}-\cos\left[\frac{\Lambda r_+}{\sqrt{2}}\right]\right)\right.\nonumber\\
&& \left. \left(\sin\left[\frac{\Lambda r_+}{\sqrt{2}}\right]
+\cos\left[\frac{\Lambda r_+}{\sqrt{2}}\right]\right)-2\left(\frac{\Lambda r_+}{\sqrt{2}}\right)^2\right]+4\left(r_+^2+a^2\right)^2\left[-\left(\frac{\Lambda r_+}{\sqrt{2}}\right)^2\right.\nonumber\\
 &+&\left. \left(\frac{\Lambda r_+}{\sqrt{2}}\right)^7\left(e^{-\frac{{\Lambda r_+}}{\sqrt{2}}}-\left(\cos\left[\frac{\Lambda r_+}{\sqrt{2}}\right]-\sin\left[\frac{\Lambda r_+}{\sqrt{2}}\right]\right)\right)  \right].
\end{eqnarray}
   
\begin{figure*}[ht]
\begin{center}$
\begin{array}{c c c c}
\includegraphics[width=.75\linewidth]{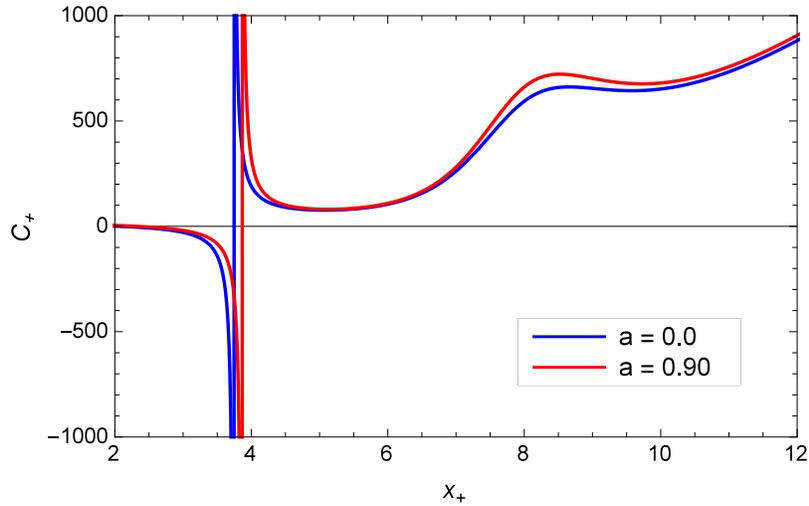}
\end{array}$
 \end{center}
\caption{The plot of specific heat $C_+$ as the function of dimensionless parameter $x=\Lambda r/\sqrt{2}$ for different values of angular momentum   $a$  with fixed value of  $\Lambda=4\sqrt{2}\pi$. }
\label{fi7}
\end{figure*}
\begin{figure*}[ht]
\begin{center}$
\begin{array}{c c c c}
\includegraphics[width=.75\linewidth]{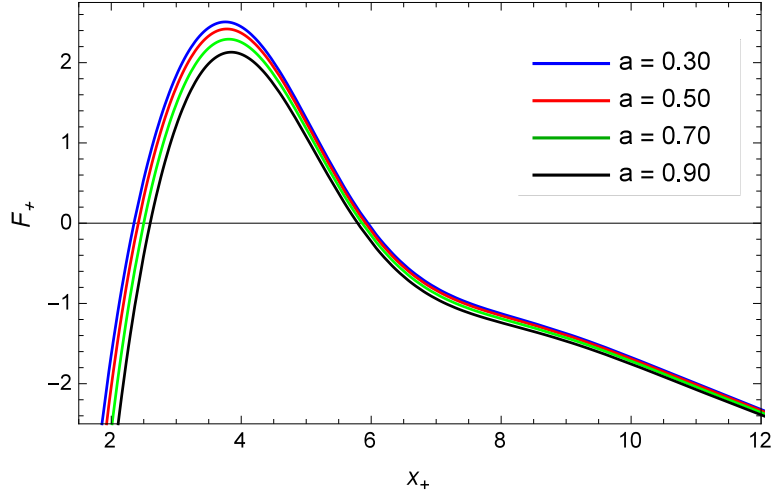}
\end{array}$
 \end{center}
\caption{The plot of free energy $C_+$ as the function of dimensionless parameter $x=\Lambda r/\sqrt{2}$ for different values of angular momentum   $a$  with fixed value of  $\Lambda=4\sqrt{2}\pi$. }
\label{fig:7}
\end{figure*}

{ In order to study the stability of rotating Lee-Wick black hole, we plot the  heat capacity with respect to dimensionless parameter  $x$ in Fig \ref{fig:7}.
From Eq. (\ref{f7}), it can be observed that the specific heat diverges at 
{$r_+=3.75$} this discontinuity of the  heat capacity represents a point of phase 
transition.}

Now, we study the behaviour of Gibbs free energy for the global stability of black hole thermodynamics. The definition of Gibbs free energy is given by
\be
G_+=M_+-T_+S_+.
\label{free}
\ee
Substituting the mass, temperature and entropy from Eqs. (\ref{mass}),  (\ref{tlw}) and (\ref{entropy})  into  (\ref{free}) we get
\begin{eqnarray}
G_+&=&\frac{\left(a^2+r_+^2\right)}{2r_+ \left(1-e^{-\frac{\Lambda r_+}{\sqrt{2}}}\left(\left(1+\frac{\Lambda r_+}{\sqrt{2}}\right)\cos\left[\frac{\Lambda r_+}{\sqrt{2}}\right]+\frac{\Lambda r_+}{\sqrt{2}}  \sin\left[\frac{\Lambda r_+}{\sqrt{2}}\right]\right)\right)}+\left(\frac{r_+^2-a^2}{4\pi(r_+^2+a^2)}\right)\nonumber\\&&\left(\frac{e^{-\frac{r_+ \Lambda }{\sqrt{2}}}\left(\left(1+ \frac{\Lambda r_+ }{\sqrt{2}}\right) \cos\left[\frac{ \Lambda r_+}{\sqrt{2}}\right]+r_+ \left( \frac{\Lambda r_+ }{\sqrt{2}}+\frac{\Lambda }{2}\left(\frac{a^2+r_+^2}{a^2-r_+^2} \right)\right) \sin\left[\frac{ \Lambda r_+}{\sqrt{2}}\right]\right)-1}{e^{-\frac{r_+ \Lambda }{\sqrt{2}}}\left(\left(1+ \frac{\Lambda r_+}{\sqrt{2}}\right) \cos\left[\frac{\Lambda r_+}{\sqrt{2}}\right]+\frac{\Lambda r_+}{\sqrt{2}}\sin\left[\frac{\Lambda r_+}{\sqrt{2}}\right]-r_+\right)}\right)\nonumber\\&&\int dr_+\frac{(a^2+r_+^2) e^{-r_+\Lambda/\sqrt{2}}}{e^{-\frac{\Lambda r_+}
{\sqrt{2}}}\left(\left(2+{\sqrt{2}}\,r_+ \Lambda\right)\cos\left[\frac{\Lambda 
r_+}{\sqrt{2}}\right]+{\sqrt{2}}  \sin\left[\frac{\Lambda r_+}{\sqrt{2}}
\right]\right)}.
\end{eqnarray}

{ We analyse the stability of the black hole by studying the nature of free energy which is plotted for the 
different values of rotation parameter $(a)$   in Fig.  \ref{fig:7}. Here we see that the free energy has a maximum value where the specific heat diverges (see Fig. \ref{fi7}) which can be identified as the extremal points of the Hawking temperature (see Fig. \ref{fig:6}). We also notice  that the rotating black hole is unstable for small values of $r_+$ and stable for  large horizon radius. It means that the large black hole is stable. }

\section{Conclusions} \label{sec4}
We have extended the investigation of static and spherically symmetric Lee-
Wick solution. In fact, we have  obtained a rotating Lee-Wick black hole 
solution by exploiting a corrected Newman-Janis algorithm. Specifically, in 
order to derive a rotational counterpart, 
we first considered   static spherically symmetric Lee-Wick black hole 
solution 
and  made a coordinate transformation  
from the Boyer-Lindquist coordinates  to the Eddington-Finkelstei 
coordinates. Moreover, we have written the contravariant components of the 
metric tensor in the advanced null Eddington-Finkelstei coordinates in    the 
null tetrad form. After that we  have performed a complex coordinate 
transformations followed by the coordinate transformation in order to   
retain the metric in Boyer-Lindquist coordinates. Since we are left with a 
transcendental equation which can not be solved analytically.  Hence, 
we have tried to do a numerical analysis by comparing the metric function $
\Delta$ equal to zero which  reveals the possibilities for finding  the non-zero values of angular momentum and  cosmological  constant.  

Also, we have discussed the basic thermodynamics properties of this rotating 
Lee-Wick black hole. In order to do so, first we have  evaluated a mass 
parameter and angular velocity. Following standard black hole chemistry from 
areal-law, we have  
derived the Hawking temperature of rotating Lee-Wick black hole  and 
expressed in terms of 
 temperature of Kerr black hole. 
 In order to study the behaviour of black hole with horizon radius, we have 
 plotted a graph and found  that the temperature of rotating Lee-Wick black 
 holes is increasing for smaller rotation parameter. We also found that  the 
 effects of rotation parameter become
more significant for relatively smaller black holes. The
stability of this black holes is studied by evaluating heat capacity of the 
black hole and  checking the sign of heat capacity. Remarkably, the system 
exhibits various phase transitions. A comparative analysis is done and 
found that for the stationary case (when rotation parameter is turned-off) 
the phase transition occurs only once for small sized black holes. In 
contrast, when rotation parameter acquires non-zero values, the phase 
transition occurred at various points. 
The graph signified that  in contrast to  the stationary case, 
 we have found a stable region for the larger rotating black holes also. 
The results of the paper are quite interesting and yields new insight for 
future investigations.

\appendix
\section{Verification of black hole solution by equation of motion}
In order to compute the energy momentum tensor, we use the following tetrad
\begin{eqnarray}
&&e^{\mu}_t=\frac{1}{\Sigma\sqrt{\Delta}}(r^2+a^2,0,0,a),\\
&&e^{\mu}_r=\frac{\sqrt{\Delta}}{\Sigma}(0,1,0,0),\\
&& e^{\mu}_{\theta}=\frac{1}{\Sigma}(0,0,1,0),\\
&&e^{\mu}_{\phi}=-\frac{1}{\Sigma\sin\theta}(r^2+a^2,0,0,a).
\end{eqnarray}
The non-zero component of energy momentum tensor are given by
\begin{eqnarray}
 &&{\tilde \rho(r)}=-{\tilde P}_r=\frac{M\Lambda^2}{4\pi r}\sin \frac{r\Lambda}{\sqrt{2}}e^{-\frac{r\Lambda}{\sqrt{2}}},\\
&&  {\tilde P}_{\theta}={\tilde P}_{\phi}=\frac{M\Lambda^2 e^{-\frac{r\Lambda}{\sqrt{2}}}}{16\pi r}\left(\sqrt{2}\cos \frac{r\Lambda}{\sqrt{2}}+(2-\sqrt{2}r)\sin \frac{r\Lambda}{\sqrt{2}}-8\sin \frac{r\Lambda}{\sqrt{2}}\frac{M-rM'}{2M-r}\right).
\end{eqnarray}
The equation of motion of the action (\ref{1}) can be written as
\begin{eqnarray}
G_t^t&=&-\frac{1}{4\Sigma^2(f(r,\theta)E(r,\theta)+B(r,\theta))^2}\Big[-E^2\,f\,\frac{\Sigma}{\Delta}\Delta'-E^2f\left(\frac{\Sigma}{\Delta}\right)'^2\Delta\nonumber\\ 
&-& EB^2\Delta'^2\frac{\Sigma}{\Delta}-E \Delta B^2\left(\frac{\Sigma}{\Delta}\right)'^2+2\Sigma\Delta E'' f E+2E\left(\frac{\Sigma}{\Delta}\right)''\Sigma B^2+2E \Delta''\Sigma B^2\nonumber\\
&+& 2E^2f \left(\frac{\Sigma}{\Delta}\right)''\Sigma+2\left(\frac{\Sigma}{\Delta}\right)^2 \Delta E''B^2+2 E^2f\Delta''\Sigma-E^2f\Delta' \left(\frac{\Sigma}{\Delta}\right)'\frac{\Sigma}{\Delta}\nonumber\\
&-& E^2f  \left(\frac{\Sigma}{\Delta}\right)'\Delta'\Delta-E \left(\frac{\Sigma}{\Delta}\right)'\Delta'\Delta B^2-E \left(\frac{\Sigma}{\Delta}\right)'\Delta'\frac{\Sigma}{\Delta}B^2+ \left(\frac{\Sigma}{\Delta}\right)\Delta B'E
\nonumber\\
&+& \left(\frac{\Sigma}{\Delta}\right)^2\Delta B'E-\left(\frac{\Sigma}{\Delta}\right)'\Delta'E'B^2+2 \Sigma\Delta E''f E+\left(\frac{\Sigma}{\Delta}\right)'\Sigma E' fE\nonumber\\
&+& \Delta'  \Sigma E' f E-2\Sigma\Delta BB'E'-2  \left(\frac{\Sigma}{\Delta}\right)^2 \Delta B B'E'- \left(\frac{\Sigma}{\Delta}\right)'\Delta^2 E'fE\nonumber\\
&-& \left(\frac{\Sigma}{\Delta}\right)' \left(\frac{\Sigma}{\Delta}\right)\Delta E'B^2-\Delta' \left(\frac{\Sigma}{\Delta}\right)^2E' f E-\Delta' \Sigma E'B^2-E'^2\Sigma\Delta f\nonumber\\
&-& \Delta' \left(\frac{\Sigma}{\Delta}\right)^2 E'B^2-E'^2 \left(\frac{\Sigma}{\Delta}\right)^2\Delta f\Big],\label{1a}
\end{eqnarray}
\begin{eqnarray}
G^t_{\phi}&=&-\frac{1}{4\Sigma^2(f(r,\theta)E(r,\theta)+B(r,\theta))^2}\Big[\Delta'^2\frac{\Sigma}{\Delta}B^3+\left( \frac{d}{d\theta}\frac{\Sigma}{\Delta}\right)^2\Delta B^3\nonumber\\
&+& E f B \left( \frac{d}{d\theta}\frac{\Sigma}{\Delta}\right)^2\Delta + E f B\Delta'^2\frac{\Sigma}{\Delta}-2E f B \left( \frac{d^2}{d\theta^2}\frac{\Sigma}{\Delta}\right)\Sigma -2B'' \Sigma \Delta f E\nonumber\\
&-& 2B''\frac{\Sigma^2}{\Delta} f E-\frac{dB}{d\theta}\frac{dC}{d\theta}\Sigma f E-  2E f B\Delta''\Sigma -B'\Delta'\Sigma f E-2Ef B \Delta'' \Sigma\nonumber\\
&-& f'\Sigma\Delta B E'+\frac{dA}{d\theta}\frac{dB}{d\theta}\frac{\Sigma^2}{\Delta}E-\frac{dA}{d\theta}\frac{dE}{d\theta} \frac{\Sigma^2}{\Delta}B+B'C'\Delta^2 f E\nonumber\\
&-&\frac{dB}{d\theta} \frac{dC}{d\theta}\Sigma B^2+E'\Sigma\Delta f B'+\frac{dB}{d\theta}\frac{d\Delta}{d\theta} \frac{\Sigma^2}{\Delta^2} f E-B'\Delta'\Sigma B^2+\frac{dE}{d\theta}\frac{dB} f \frac{\Sigma^2}{\Delta}{d\theta}\nonumber\\
&-&2B''\Sigma\Delta B^2-2\frac{d^2B}{d\theta^2} B^2\frac{\Sigma^2}{\Delta}+E f BC'\Delta'\Delta+ E f B\frac{dC}{d\theta} \frac{d\Delta}{d\theta}\frac{\Sigma}{\Delta}-2\Delta''\Sigma B^3 \nonumber\\
&+&C'\Delta'\Delta B^3+ B \Sigma\Delta B''+B\frac{\Sigma^2}{\Delta} \left(\frac{dB}{d\theta}\right)^2+B'C'\Delta^2B^2+\frac{dB}{d\theta}\frac{d\Delta}{d\theta}\frac{\Sigma^2}{\Delta^2}\nonumber\\
&+&A'\Sigma\Delta E B'-2\frac{d^C}{d\theta^2}\Delta B^3\Big],\label{2a}
\end{eqnarray}
\begin{eqnarray}
G^r_r&=&\frac{1}{4\Sigma^2(f(r,\theta)E(r,\theta)+B(r,\theta))^2}\Big[\Delta^2B'^2B^2+2\Delta'\Delta B^3B'-\Sigma B^2\left(\frac{dB}{d\theta}\right)^2\nonumber\\&-&\Sigma E^2\left(\frac{dA}{d\theta}\right)^2-2\left(\frac{d\Delta}{d\theta}\right) \frac{\Sigma}{\Delta} B^3\left(\frac{dB}{d\theta}\right)-\Sigma f^2\left(\frac{dE}{d\theta}\right)^2+\Delta^2B'^2 f E\nonumber\\&+&\Delta^2 E'f'B^2+2\Delta'\Delta B B'f E+\Sigma\left(\frac{dE}{d\theta}\right)\left(\frac{df}{d\theta}\right) f E-4\Sigma E B \left(\frac{df}{d\theta}\right)\left(\frac{dB}{d\theta}\right)\nonumber\\&-&4\Sigma B f \left(\frac{dB}{d\theta}\right)\left(\frac{dE}{d\theta}\right)-2\left(\frac{d\Delta}{d\theta}\right)\left(\frac{dB}{d\theta}\right) \frac{\Sigma}{\Delta}B f E+3\Sigma \left(\frac{dE}{d\theta}\right)\left(\frac{df}{d\theta}\right) B^2\nonumber\\&+& 3\Sigma\left(\frac{dE}{d\theta}\right)\left(\frac{dA}{d\theta}\right) B^2+3\Sigma\left(\frac{dB}{d\theta}\right)^2 f E-\left(\frac{d\Delta}{d\theta}\right)\frac{\Sigma}{\Delta}E^2 f \left(\frac{df}{d\theta}\right)\nonumber\\&-&\left(\frac{d\Delta}{d\theta}\right)\frac{\Sigma}{\Delta} E B^2\left(\frac{dA}{d\theta}\right)-\left(\frac{d\Delta}{d\theta}\right)\frac{\Sigma}{\Delta} A^2 E \left(\frac{dE}{d\theta}\right)-\left(\frac{d\Delta}{d\theta}\right)\frac{\Sigma}{\Delta}f B^2\left(\frac{dE}{d\theta}\right)\nonumber\\&+&\Delta^2 E' f' f E+\Delta'\Delta E^2f'f+\Delta' \Delta E f'B^2+\Delta' \Delta f E' E+\Delta'\Delta f E' B^2\nonumber\\&+&4\Sigma B \left(\frac{d B}{d\theta^2}\right) fE+4 \Sigma E^2\left(\frac{d^2f}{d\theta^2}\right)f+2\Sigma f B^2\left(\frac{d^E}{d\theta^2}\right)+2\Sigma f^2 \left(\frac{d^2E}{d\theta^2}\right)E\nonumber\\&+&2\Sigma E \left(\frac{d^2A}{d\theta^2}\right)B^2+4 \Sigma B^2 \left(\frac{d^2B}{d\theta^2}\right)\Big],\label{3a}
\end{eqnarray}
\begin{eqnarray}
G_{\theta}^{\theta}&=&\frac{1}{4\Sigma^2(f(r,\theta)E(r,\theta)+B(r,\theta))^2}\Big[\Sigma B'^2B^2-\Sigma E^2A'^2-2C'\Delta B^3B'-\Sigma f^2 E'^2\nonumber\\&+&2\left(\frac{d}{d\theta}\frac{\Sigma}{\Delta}\right)\frac{\Sigma}{\Delta}B^3\left(\frac{dB}{d\theta}\right)+\left(\frac{dE}{d\theta}\right)\left(\frac{df}{d\theta}\right)B^2\left(\frac{\Sigma}{\Delta}\right)^2\left(\frac{dB}{d\theta}\right)^2 f E\nonumber\\&+&\left(\frac{\Sigma}{\Delta}\right)BfE\left(\frac{dB}{d\theta}\right)\left(\frac{dC}{d\theta}\right)+\Sigma E'f' f E-4\Sigma E f' B B'-4\Sigma B B' f E'\nonumber\\&-&2\left(\frac{\Sigma}{\Delta}\right)'\Delta B B' f E+\left(\frac{d}{d\theta}\frac{\Sigma}{\Delta}\right)\frac{\Sigma}{\Delta} E^2\frac{df}{d\theta}f+\left(\frac{d}{d\theta}\frac{\Sigma}{\Delta}\right)\frac{\Sigma}{\Delta} E\frac{df}{d\theta}B^2\nonumber\\&+&\left(\frac{d}{d\theta}\frac{\Sigma}{\Delta}\right)\frac{\Sigma}{\Delta} A^2\frac{dE}{d\theta}E+\left(\frac{d}{d\theta}\frac{\Sigma}{\Delta}\right)\frac{\Sigma}{\Delta} \frac{dE}{d\theta}fB^2+\left(\frac{\Sigma}{\Delta}\right)^2\frac{dE}{d\theta}\frac{df}{d\theta}+3\Sigma E' f' B^2\nonumber\\&+& 3\Sigma B'^2f E-\left(\frac{\Sigma}{\Delta}\right)'\Delta E^2 f f'-\left(\frac{\Sigma}{\Delta}\right)'\Delta E f'B^2-\left(\frac{\Sigma}{\Delta}\right)'\Delta  f'^2 EE'\nonumber\\&-&\left(\frac{\Sigma}{\Delta}\right)'\Delta f E'B^2+2\Sigma f E'' B^2+2\Sigma f^2E'' E+2\Sigma E^2 f'' f\nonumber\\&+&4 \Sigma B^3B''+4\Sigma BB'' f E+\frac{\Sigma^2}{\Delta^2} B^2\left(\frac{dB}{d\theta}\right)^2\Big],\label{4a}
\end{eqnarray}
\begin{eqnarray}
G_{\phi}^{\phi}&=&\frac{1}{4\Sigma^2(f(r,\theta)E(r,\theta)+B(r,\theta))^2}\Big[2 \Delta''\Sigma fB^2+2E\left(\frac{d^2}{d\theta^2}\frac{\Sigma}{\Delta}\right)f^2+2E\Delta''\Sigma f^2\nonumber\\
&-&\Delta'^2\frac{\Sigma}{\Delta}f B^2-\left(\frac{d}{d\theta}\frac{\Sigma}{\Delta}\right)^2\Delta f B^2-E\Delta'^2 \frac{\Sigma}{\Delta}f^2-E\left(\frac{d}{d\theta}\frac{\Sigma}{\Delta}\right)^2\Delta f^2\nonumber\\
&-&f'\left(\frac{\Sigma}{\Delta}\right)' \Delta^2 f E-2f' \Sigma\Delta B B'+\frac{df}{d\theta}\left(\frac{d}{d\theta}\frac{\Sigma}{\Delta}\right) \Sigma B^2-\frac{df}{d\theta}\frac{d\Delta}{d\theta}f \left(\frac{\Sigma}{\Delta}\right)^2 \frac{dB}{d\theta}\nonumber\\
&+&\frac{df}{d\theta} \left(\frac{\Sigma}{\Delta}\right)^2\Delta B^2+f' \Delta' \Sigma B^2+2  \left(\frac{\Sigma}{\Delta}\right)^2\frac{d^2f}{d\theta^2}\Delta B^2+2 \frac{d^2f}{d\theta^2}\Sigma\Delta B^2+2f''\Sigma \Delta f B^2\nonumber\\
&+&2\left(\frac{d^2}{d\theta^2}\frac{\Sigma}{\Delta}\right)\Sigma f B^2+\frac{d^2f}{d\theta^2}\frac{\Sigma^2}{\delta} f E+f''\Sigma \Delta f E+\frac{df}{d\theta}\left(\frac{d}{d\theta}\frac{\Sigma}{\Delta}\right)\Sigma f E+f'\Delta'\Sigma f E\nonumber\\
&-&f'\left(\frac{\Sigma}{\Delta}\right)'\Delta^2B^2-f'^2 \Sigma \Delta E+B'^2\Sigma \Delta f- \frac{df}{d\theta}\frac{d\Delta}{d\theta}\left(\frac{\Sigma}{\Delta}\right)^2 B^2-\left(\frac{d}{d\theta}\frac{\Sigma}{\Delta}\right)\frac{d\Delta}{d\theta}\frac{\Sigma}{\Delta} fB^2\nonumber\\
&-&\left(\frac{\Sigma}{\Delta}\right)'\Delta'\Delta f B^2-E\frac{d\Delta}{d\theta}\frac{\Sigma}{\Delta}f^2-E\left(\frac{\Sigma}{\Delta}\right)'\Delta'\Delta f^2-\left(\frac{df}{d\theta}\right)^2\frac{\Sigma^2}{\Delta}E+\nonumber\\
&&\left(\frac{dB}{d\theta}\right)^2\frac{\Sigma^2}{\Delta}f\Big],\label{5a}
\end{eqnarray}
where
\begin{eqnarray}
&&f=1-\frac{2rm(r)}{\Sigma},\\
&&B=\frac{r\, a \,m(r)\sin^2\theta}{\Sigma},\\
&&\Sigma=r^2+a^2\,\cos^2\theta,\\
&&\Delta=r^2-2Mr+a^2+2Me^{-\frac{\Lambda r}{\sqrt{2}}}\left[\left(1+\frac{\Lambda r}{\sqrt{2}}\right)\cos\frac{\Lambda r}{\sqrt{2}}+\sin\frac{\Lambda r}{\sqrt{2}}\right],\\
&&E=r^2+a^2+\frac{2m(r)a^2\sin^2\theta}{\Sigma}.
\end{eqnarray}
By substituting the above value of  $f, B,\Sigma, \Delta$  {and} $E$  in Eqs. (\ref{1a}), (\ref{2a}), (\ref{3a}), (\ref{4a}) and (\ref{5a}) and then substituting the values of $G_{tt}, G_{rr}, G_{\theta\theta} \,\text{and} \,G_{\phi\phi}$ with  $T_{tt}, T_{rr}, T_{\theta\theta}$  {and}  $T_{\phi\phi}$ in Eq. (\ref{3}),   Einstein field equation gets satisfied.
\section{Black Hole Entropy}

Next, let us focus to an important thermodynamic quantity,
namely, entropy $S_+$ of the black hole in term of  horizon radius. The entropy of the  rotating black hole  black hole  is calculated  by using the  Wald entropy relation \cite{wald} as 
\be
S_+=\int\frac{\delta I}{\delta R_{ab\alpha\beta}}\epsilon^{a\alpha}\epsilon^{b\beta}\sqrt{h}d^2\Omega_2 ,\qquad \text{with}\qquad I=\int d^4x\sqrt{-g}\mathcal{L},\label{b1}
\ee
where $S_+ $, $I$, $ \epsilon^{ab} $, $ h $ and $ R_{ab\alpha\beta}  $ are the entropy,   action,  bi-normal to the horizon,  induced metric on the horizon and the Riemann tensor, respectively. To calculate the entropy an antisymmetric second rank tensor $ \epsilon_{a b} $ is constructed so that $ \epsilon_{rt}= \epsilon_{tr} =1 $.
The specific form of Lagrangian and its derivative with respect to 
Riemann tensor are given by
\be
\mathcal{L}=\frac{1}{16\pi}R_{a b\alpha\beta}g^{b\alpha}g^{a\beta} \qquad \mbox{and} \qquad  \frac{\partial\mathcal{L}}{\partial R_{a b\alpha\beta}}=\frac{1}{16\pi}\frac{1}{2}(g^{a\alpha}g^{b\beta}-g^{b\alpha}g^{a\beta}).
\ee
The definition (\ref{b1}) corresponding to the above expressions  leads to the Wald entropy as 
\begin{eqnarray}
&&S_+=\frac{1}{8}\int\frac{1}{2}(g^{a\alpha}g^{b\beta}-g^{b\alpha}g^{a\beta})\epsilon_{a b}\epsilon^{\alpha\beta}\sqrt{h}d\theta d\phi,\nonumber\\
&&S_+=\frac{1}{8}g^{tt}g^{rr}\,\sqrt{h}\,d\theta d\phi=\frac{1}{4}\int r^2\,d\theta d\phi,\nonumber\\
&&S_+=\pi r_+^2=\frac{A}{4}.
\label{entropy1}
\end{eqnarray}
This expression of entropy resembles with the standard Bekenstein-Hawking area-law.  


\begin{thebibliography}{99}


\bibitem{1} R. Utiyama and B.S. DeWitt,  J. Math. Phys. 3
(1962) 608.
\bibitem{2}M. Asorey, J. L. Lopez, and I.L. Shapiro, Int. J. Mod. Phys. A 12, 5711 (1997).
\bibitem{3} L. Modesto and I.L. Shapiro,   Phys. Lett. B 755 (2016) 279.
\bibitem{4} L. Modesto,   Nucl. Phys. B 909 (2016) 584.

\bibitem{w1}
T. D. Lee and G. C. Wick, Nucl. Phys. B  {\bf 9}, 209 (1969).
\bibitem{w2}
T. D. Lee and G. C. Wick, Phys. Rev. D {\bf 2}, 1033 (1970).
\bibitem{w3}
R. E. Cutkosky, P. V. Landshoff, D. I. Olive and J. C. Polkinghorne, Nucl. Phys. B {\bf 12}, 281 (1969). 
\bibitem{l1}
L. Modesto, Nucl. Phys. B 909, 584 (2016) [arXiv:1602.02421 [hep-th]].
\bibitem{l2}
 B. L. Giacchini, arXiv:1609.05432 [hep-th].
\bibitem{l3}
 G. P. de Brito, P. I. C. Caneda, Y. M. P. Gomes, J. T. G. Junior and V. Nikoofard, arXiv:1610.01480 [hep-th].
\bibitem{l4}
A. Accioly, B. L. Giacchini and I. L. Shapiro, arXiv:1610.05260 [gr-qc].
\bibitem{5}
I. L. Shapiro, Phys. Lett. B 744, 67 (2015).
\bibitem{6}
 L. Modesto and I. L. Shapiro, Phys. Lett. B 755, 279 (2016).
\bibitem{7}
T. D. Lee and G. C. Wick, Nucl. Phys. B 9, 209 (1969).
\bibitem{8}
 T. D. Lee and G. C. Wick, Phys. Rev. D 2, 1033 (1970).
\bibitem{9}
R. E. Cutkosky, P. V. Landshoff, D. I. Olive and J. C. Polkinghorne, Nucl. Phys. B 12, 281 (1969).
\bibitem{Bambi:2016wmo}   C.~Bambi, L.~Modesto and Y.~Wang,    Phys. Lett.  B { 764}, 306 (2017).

\bibitem{10}
 S. Carlip, Int. J. Mod. Phys. D. 23, 1430023 (2014).
\bibitem{11}
A. Bekenstein, Nuovo Cimento Lett. 4, 99 (1972).
\bibitem{12}
 S. W Hawking, Comm. in Math. Phys. 43, 199 (1975)
\bibitem{13}
J. D.  Bekenstein,  Physical Review D. 9, 3292 (1974).
\bibitem{14}
R. M Wald,  Living Reviews in Relativity. 4,1  (2001)
\bibitem{haw}
S. W . Hawking and D. N. Page, Commun, Math. Phys. {  87}, 577  (1983).
\bibitem{Newman:1965tw}
E.~Newman and A.~Janis, 
J.\ Math.\ Phys.\   {6}, 915 (1965).

\bibitem{Hansen:2013owa} 
D.~Hansen and N.~Yunes, 
Phys.\ Rev.\ D {  88}, 104020 (2013).

\bibitem{az}M. Azreg-Aınou,   Phys. Rev. D 90 (2014) 064041.
 \bibitem{ze}B. Toshmatov,  Z. Stuchlík, B. Ahmedov, Eur. Phys. J. Plus 132 (2017) 2.
\bibitem{Ahmed:2020dzj}
F.~Ahmed, D.~V.~Singh and S.~G.~Ghosh, 
 arXiv:2008.10241 [gr-qc].
\bibitem{Ahmed:2020ifa}
F.~Ahmed, D.~V.~Singh and S.~G.~Ghosh, arXiv:2002.12031 [gr-qc].
\bibitem{Ali:2019rjn}
M.~S.~Ali and S.~G.~Ghosh,  arXiv:1906.11284 [gr-qc].
\bibitem{Ali:2019myr}
M.~S.~Ali and S.~G.~Ghosh, Phys. Rev. D  {99} (2019)  024015.
\bibitem{Singh:2017bwj}
D.~V.~Singh, M.~S.~Ali and S.~G.~Ghosh,
Int. J. Mod. Phys. D \textbf{27} (2018) no.12, 1850108.
\bibitem{Chaturvedi:2016fea}
P.~Chaturvedi, N.~K.~Singh and D.~V.~Singh,
Int. J. Mod. Phys. D \textbf{26} (2017) no.08, 1750082.
\bibitem{wald}
R. M. Wald, Phys. Rev. D {\bf 43}, R3427 (1993).





  
 
 

\end{thebibliography}
\end{document}